# Double-edged Role of Interactions in Superconducting Twisted Bilayer Graphene

Xueshi Gao[1], Alejandro Jimeno-Pozo[2], Pierre A. Pantaleon[2], Emilio Codecido[1,3], Daria L. Sharifi[1], Zheneng Zhang[1], Youwei Liu[1], Kenji Watanabe[4], Takashi Taniguchi[5], Marc W. Bockrath[1], Francisco Guinea[2,6], Chun Ning Lau[1*]

[1] Department of Physics, The Ohio State University, Columbus, OH 43221, USA
[2] Imdea Nanoscience, Faraday 9, 28049 Madrid, Spain
[3] Lakeshore Cryotronics, 575 McCorkle Blvd, Westerville, Ohio 43082, USA
[4] Research Center for Electronic and Optical Materials, National Institute for Materials Science, 1-1 Namiki, Tsukuba 305-0044, Japan
[5] Research Center for Materials Nanoarchitectonics, National Institute for Materials Science, 1-1 Namiki, Tsukuba 305-0044, Japan
[6] Donostia International Physics Center, Manuel Lardizabal 4, 20018 Donostia-San Sebastian, Spain

Abstract

**For the unconventional superconducting phases in moiré materials, a critical question is the role played by electronic interactions in the formation of Cooper pairs. In twisted bilayer graphene (tBLG), the strength of electronic interactions can be reduced by increasing the twist angle $\theta$ or screening provided by the dielectric medium. In this work, we place tBLG at 3-4 nm above bulk SrTiO$_3$ substrates, which have a large yet tunable dielectric constant $\varepsilon$. By raising $\varepsilon$ in situ in a magic angle device, we observe suppression of both the height and the width of the entire superconducting dome, thus demonstrating that, unlike conventional superconductors, the pairing mechanism in tBLG is strongly dependent on electronic interactions. Interestingly, in contrast to the absence of superconductivity in devices on SiO$_2$ with $\theta>1.3°$, we observe a superconducting pocket in a large-angle ($\theta=1.4°$) tBLG/STO device while the correlated insulating states are absent. These experimental results are in qualitative agreement with a theoretical model in which the pairing mechanism arises from Coulomb interactions that are screened by plasmons, electron-hole pairs, and longitudinal acoustic phonons. Our results highlight the unconventional nature of the superconductivity in tBLG, the double-edged role played by electronic interactions in its formation, as well as their complex interplay with the correlated insulating states.**

The discovery of superconductivity (SC) and correlated insulating (CI) states in twisted bilayer[1-3] graphene has ushered in a gold rush in moiré materials[4, 5], yet the rapid pace of discovery is paralleled by the number of questions that remain outstanding. One of the most fundamental questions is the pairing mechanism of superconductivity in tBLG. Various mechanisms have been proposed, e.g. electron-electron interactions, excitonic insulators, the van Hove singularity, or electron-phonon coupling, and skymions[6-23]. Experimentally, one can shed light on this question by tuning the strength of Coulomb interactions, which is controlled by two parameters: the onsite interaction $U \sim e^2/(4\pi\varepsilon a_M)$, and the bandwidth $W$, which is sensitively dependent on the twist angle $\theta$, expecting to vanish at the magic angle $\theta \sim 1.05°$[2]. Here $e$ is the electron charge, $a_M$ is the moiré wavelength, and $\varepsilon$ is the dielectric constant.

---

* Email: lau.232@osu.edu

Prior experiments [24, 25] showed that, when a metallic gate is placed ~7-12 nm away from tBLG, or when the twist angle $\theta$ is detuned from the magic angle of ~1.05°, the CI states are suppressed while SC remains robust. The superconducting state's seeming insensitivity to interaction strength is often construed as evidence for phonon-driven pairing. However, these experiments are performed by using samples with varying twist angles, disorder, mobilities, and strains, thus direct comparison of different devices is difficult. In another experiment, the SC transition temperature $T_c$ is slightly suppressed when the Fermi level of a nearby Bernal-stacked BLG is tuned into its band gap[26]. This result appears to suggest that SC is enhanced by weakened electronic interactions, though such interpretation is complicated by the role of noise filtering, and the change in $T_c$ was minute. On the theoretical front, a recent work pointed out that, for plasmon-driven superconductivity, substrate-provided screening becomes significant only when the metallic layer is placed at less than 3nm from tBLG[20]. Thus, the pairing mechanism of SC in tBLG remains a subject of debate, despite much efforts.

In this work, by *in situ* tuning of $U$ in tBLG devices on SrTiO$_3$ (STO) substrates, we demonstrate that screening electronic interactions is a double-edged sword for the emergence of SC in tBLG. At the magic angle *($\theta \sim 1.05°$)*, we observe complete tuning of the entire SC dome, and dramatic suppression of SC when the substrate's effective dielectric constant increases, providing strong evidence that that the pairing process is partly or wholly driven by electronic interactions. Paradoxically and surprisingly, in a tBLG/STO device that is detuned significantly from the magic angle ($\theta=1.4°$) and hence have large $W$, SC is observed while CIs are absent at zero magnetic field, indicating the competing natures of these two phases. The suppression of SC by enhanced screening as well as the observation of SC in large-angle tBLG are understood qualitatively in terms of interaction-driven pairing. Our results shed light on the pairing mechanism and opening venues for understanding SC and competing phases in two-dimensional (2D) moiré system.

We fabricate hexagonal BN (hBN)-encapsulated tBLG heterostructures, which are deposited on 500 μm-thick STO substrates that serve as back gate dielectrics (Fig 1A). STO is an incipient paraelectric, with a temperature-dependent dielectric constant that reaches 25,000 at low temperatures[27]. Thickness of the bottom hBN layer is chosen to be $d \sim 3-4$ nm, which is sufficiently thin to maximize the effect of the large dielectric constant of STO, but still thick enough to enable high device mobility. Applying a voltage to the STO back gate controls the polarization phase of the STO substrate, hence the effective dielectric environment (see Supplementary Materials (SM)). Each tBLG device is also coupled to a graphite top gate, which allows for controlling the total charge carrier density $n$ independent of back gate voltage $V_{bg}$. Fig. 1B displays the longitudinal resistance $R_{xx}$ at $T=1.7$K in a pumped He$^4$ cryostat for device D1, which has $\theta=1.05°$ and hBN spacer thickness $d=3$ nm, as a function of out-of-plane magnetic field $B$ and top gate voltage $V_{tg}$. At $B=0$, band insulators (BIs) are observed at $\nu=0$ and 4 ($\nu=-4$ is out of range), and CI states in $R_{xx}$ are observed at $\nu=1, \pm2$, and 3. Here $\nu$ is the filling factor, corresponding to the number of charges per moiré unit cell (Fig. 1C). Landau fans originate from most of the all integer fillings with different broken symmetries (Fig. 1B, 1D), in agreement with prior works on tBLG, and underscoring the high quality of the devices despite the additional long range scatterers introduced by the STO substrate.

We then measure the device in a dilution refrigerator at base temperature of 20 mK. Fig. 1E displays $R_{xx}$ vs $V_{tg}$ and $V_{bg}$ at $B=0$, and Fig. 1F sketches the corresponding BI and CI phases at integer fillings, as well as the superconducting phase observed close to ν = -2. Fig. 1G-H plots line traces $R_{xx}$ (red, left axis) and $R_{xy}$ (blue, right axis) vs $V_{tg}$ at two different values of $V_{bg}$. In particular, the Hall resistance data enables measurements of total charge density $n$ and hence determination of ν.

Strikingly, unlike standard dual-gated graphene devices on Si/SiO$_2$, where features at iso-carrier-density occur along diagonal straight lines on the $V_{tg}$-$V_{bg}$ plane, those in Fig. 1E are concave up, indicating that the ratio of the two gates' dielectric constants is not a constant, but changes as $V_{tg}$ and $V_{bg}$ are modulated. Since the coupling efficiency of the hBN/graphite top gate is constant (as verified by the $R_{xy}$ data), these curved features signify that the effective dielectric constant $\varepsilon_{eff}$ of the back gate is smaller at more positive $V_{bg}$ values. Such a field-dependent $\varepsilon_{eff}$ is not surprising for STO, in which polarization domains are sensitive to applied electric fields, and has indeed been observed experimentally[28, 29]. By tracing the trajectory of the device's overall charge neutrality point (CNP) on the $V_{tg}$-$V_{bg}$ plane, we estimate that $\varepsilon_{eff}$ varies from 8,000 to 25,000 in this device (see SM for details); these values are in good agreement with prior studies of STO[27].

The *in situ* tunability of $\varepsilon_{eff}$ thus offers an ideal platform to investigate the effect of screening and electronic interactions on tBLG. To this end, we characterize the superconducting dome's dependence on $T$ and $B$ at $V_{bg}$=15V, 10V, 5V, 0V and -5V, respectively. The color plots of Fig. 2A-C exhibit the normalized resistance $R_{xx}/R_N$ (where $R_N$ is the normal state resistance) as a function of $V_{tg}$ and $T$; the solid curves therein denote $T_c$, which is taken to be the temperature or $B$ values at which $R_{xx}/R_N$=20%. The extracted $T_c$ and critical magnetic field $B_c$ vs $V_{tg}$ at different $V_{bg}$ are shown in Fig. 2D and 2E, respectively. Evidently, both the height and the width of the entire superconducting dome are highly tunable by $V_{bg}$. For instance, at $V_{bg}$=15V, the optimal $T_c$ is ~0.8K, the maximum $B_c$ is 28 mT, and SC persists over 0.5V in $V_{tg}$. Successive down-stepping $V_{bg}$ weakens the observed superconducting phase: at $V_{tg}$=-5V, the optimal $T_c$ drops to ~0.2K, optimal $B_c$ is close to 0, whereas SC is restricted to a tiny range (<0.1 V in $V_{tg}$).

Such a strongly tunable superconducting dome is hitherto unreported. To determine the underlying mechanism, we note that varying $V_{bg}$ modulates two variables: the dielectric constant $\varepsilon_{eff}$ (as discussed above), and the out-of-plane displacement field $D$. In principle, a sufficiently large $D$ can overcome the strong interlayer hybridization, hence destroying the moiré potential and SC. However, the required $D$ value is expected to be unphysically high, $>\sim \hbar v_F k_\vartheta$ ~3 V/nm, where $\hbar$ is the reduced Planck constant, $v_F$=10$^6$ is the Fermi velocity of monolayer graphene, $k_\vartheta=\frac{4\pi}{3a}\theta$ is the momentum separation of the Dirac cones from the two graphene layers, and $a$=0.246 nm the lattice constant of graphene [2]. This is corroborated by a prior experimental work, in which $T_c$ exhibits no dependence on $D$ up to ±0.6 V/nm. In our experiment, $D$ induced by accessible gate voltages is estimated to be relatively modest, ~-0.2 to 0.6 V/nm. Thus we exclude the displacement field as the underlying cause for suppression of SC, and attribute the SC tuning to the modulating dielectric constant and hence screening of Coulomb interactions. As shown in Fig. 2D-E, the maximum $T_c$ and $B_c$ are suppressed a factor of 4 and 10, respectively, when $\varepsilon_{eff}$ increases from 8,000 to 19,000. We note that this behavior is opposite to that reported in ref. [26], which can be explained by the fact that screening provided by the 500 μm-thick STO is much stronger than that by a lightly doped atomic bilayer; this is also supported by the much greater extent of SC suppression in our experiment, comparing to the minute change in $T_c$ (~10 mK) in ref. [26].

Our demonstration of dramatic suppression of superconductivity with increasing screening strongly suggests that electronic interactions are important for the emergence of SC in tBLG. Electron-electron interactions modify significantly the band structure of tBLG, and their role on the superconductivity has been extensively discussed. Internal screening[15] and screening by plasmons[7, 18-20] or electron-hole pairs[21] have been predicted to produce attractive interactions and facilitate pair formation. Our results are consistent with a number of these works. It is worth noting, that, even when phonons are taken into account, as longitudinal acoustic phonons modify dielectric screening, reducing $\varepsilon_{eff}$ is expected to lead to increasing $T_c$[7]. In the case of pairing associated to the role played by plasmons, electron-hole pairs, and longitudinal acoustic phonons in the screening of electron-electron interactions, modifications provided by a nearby layer is predicted to be significant only when the layer is within ~3 nm of the tBLG, which is consistent with our device geometry[20]. On the other hand, we caution that pairing based on electron-phonon coupling only cannot be definitively ruled-out, as it may aid the pair formation in addition to the electronic interactions, and its strength may also be affected by screening[7, 22, 23].

To glimpse further into the role of interactions in superconducting tBLG, we fabricate a tBLG device D2 on STO that is detuned from the magic angle ($\theta$=1.4°, $d$=4.3 nm) and has much larger bandwidth. For devices on hBN with large-than-magic twist angles, the much larger bandwidth reduces the effect of electronic interactions at large angles. Thence, for $\theta$>1.33°, no SC has been observed in tBLG on hBN ambient pressure; in contrast, the CI states vanish at integer filling factors[24, 25], but appear robustly at fractional fillings of $\nu$=±8/3 and ±4/3 with triple unit cell reconstructions, arising from interaction-induced orbital geometric frustrations[30].

Fig. 3A displays D2's magneto-transport data $R_{xx}(V_{tg},B)$, with 4-fold degenerate Landau fans emanating from $\nu$=0 and 4. In contrast to large-$\theta$ devices on SiO$_2$, not a single CI state is observed at $B$=0, thus the large screening provided by STO suppresses its formation at integer *and* fractional fillings (though a modest resistive peak develops near $\nu$=8/3 at higher magnetic fields). Fig. 3B displays $R_{xx}$ data from a dual-gate sweep at $B$=0, where the band insulators at $\nu$=0 and ±4 are observed. Unexpectedly, a small superconducting region at $\nu$~-2.5 emerges, as shown by the encircled dark blue stripe. The superconducting nature of this region is further confirmed by the sharp peaks in $dV/dI$ curves as a function of $I$, and their suppression by a small magnetic field and increasing temperature (Fig. 3C-D).

Similar to D1, this superconducting region becomes narrower and eventually disappears as $V_{bg}$ decreases. We extract the superconducting domes $T_c(V_{tg})$ at $V_{bg}$=15V, 11V, and 7V, respectively) (Fig. 4A), and plot the optimal $T_c$ vs estimated $\varepsilon_{eff}$ (Fig. 4B). The same trend as D1 is recovered: increasing $\varepsilon_{eff}$ and screening leads to a reduction of $T_c$. We note that over this region of SC, $D$ varies from -0.04 to 0.51 V/nm, which is not expected to impact SC.

The observation of SC in this large angle device on STO is surprising, as electronic interactions are now doubly weakened by the close proximity of a high-k medium, and by the larger bandwidth. We hypothesize that the SC and CI are competing phases, as proposed previously[24, 25], but the CI states are far more sensitive to screening. As the proximity of STO quenches all CI states at integer or fractional filling, SC is then allowed to emerge.

To account for the observations, here we model the superconducting state by assuming that the pairing mechanism arises from the screened Coulomb interactions, where screening is due to plasmons, electron-hole pairs, and longitudinal acoustic phonons. The model is based on the diagrammatic Kohn-Luttinger approach. The calculation only requires parameters well-determined in decades of graphite and graphene research, and the resulting trends in the value of the superconducting critical temperature reproduce well the observations in a variety of twisted and untwisted graphene stacks[7, 31-35] (also see SM). Interestingly, we find a low, but finite, critical temperature for $\theta = 1.45^0, T_c \approx 25$mK. Results for the dependence of $T_c$ on the screening induced by the substrate are discussed in SM. The model includes longitudinal acoustic phonons, but their effect is screened both by their coupling to plasmons and electron-hole pairs, and by the presence of the substrate. The pairing mechanism discussed here, based on the repulsive Coulomb interaction, implies that $T_c \rightarrow 0$ as $\epsilon \rightarrow \infty$. While this model does not capture the quantitative aspects of the data, it does show the importance of screening in relation the changes in $T_c$, in qualitative agreement to the experiment. Future work will be required to fully understand the precise balancing of factors that induce the superconducting pairing in this system.

To sum up, our experiments demonstrated that screening is a double-edged sword for SC in tBLG: at the magic angle, it decreases electronic interactions and suppresses SC; at large twist angles, it suppresses CI states and allows SC to emerge. The surprising emergence of superconductivity in large-angle tBLG is qualitatively supported by our theoretical calculations. Our experimental data and theoretical calculations strongly suggest that electronic interactions are important for the pairing mechanism, hence SC in tBLG is unconventional. Moreover, onsite interactions, screening and bandwidth are all important for the complex interplay between different phases in tBLG. In the future, many phenomena and open questions of moiré physics await exploration using similar ε-tuning techniques. For instance, the observation of SC in large angle devices is encouraging from the theoretical perspective, since computation of their band structures is much more straightforward and less controversial than those at the magic angle; thus, detailed studies of SC and CI as a function of $\theta$, $d$ and $\varepsilon_{eff}$ promise to reveal their natures and the SC pairing mechanism. Such ε-tuning technique may also be applied to investigate correlated states in other systems, such as moiré transition metal dichalcogenides and rhombohedral-stacked few-layer graphene.

**Acknowledgement**
We thank Brian Skinner, Xiaoshan Xu, and Xin Li for helpful discussions. The experiments are supported by DOE BES Division under grant no. DE-SC0020187. Devices are fabricated using the nanofabrication facility that is supported by NSF Materials Research Science and Engineering Center Grant DMR-2011876. K.W. and T.T. acknowledge support from the JSPS KAKENHI (Grant Numbers 21H05233 and 23H02052) and World Premier International Research Center Initiative (WPI), MEXT, Japan. A.J.P, P.A.P. and F.G. acknowledge support from from the 'Severo Ochoa' Programme for Centres of Excellence in R&D (CEX2020-001039-S/AEI/10.13039/501100011033), and from NOVMOMAT, project PID2022-142162NB-I00 funded by MICIU/AEI/10.13039/501100011033 and by FEDER, UE as well as financial support through the (MAD2D-CM)-MRR MATERIALES AVANZADOS-IMDEA-NC.

Fig. 1. Device schematics and magnetotransport data from device D1 ($\theta$=1.05°, $d$=3 nm). (A). Device schematics. (B-C). $R_{xx}(V_{tg}, B)$ in kΩ and line trace $R_{xx}(V_{tg})$ at $B$=0. Data taken at $T$=1.7K. (D). Wannier diagram outlining features in (B) and their corresponding Chern numbers. Orange and blue lines indicate features with zero and non-zero Chern numbers, respectively. (E). $R_{xx}(V_{tg}, V_{bg})$ in kΩ at $T$=20 mK. (F). Schematics of the phases in (E). (G-H). $R_{xx}(V_{tg})$ and $R_{xy}(V_{tg})$ along the dotted lines in (E).

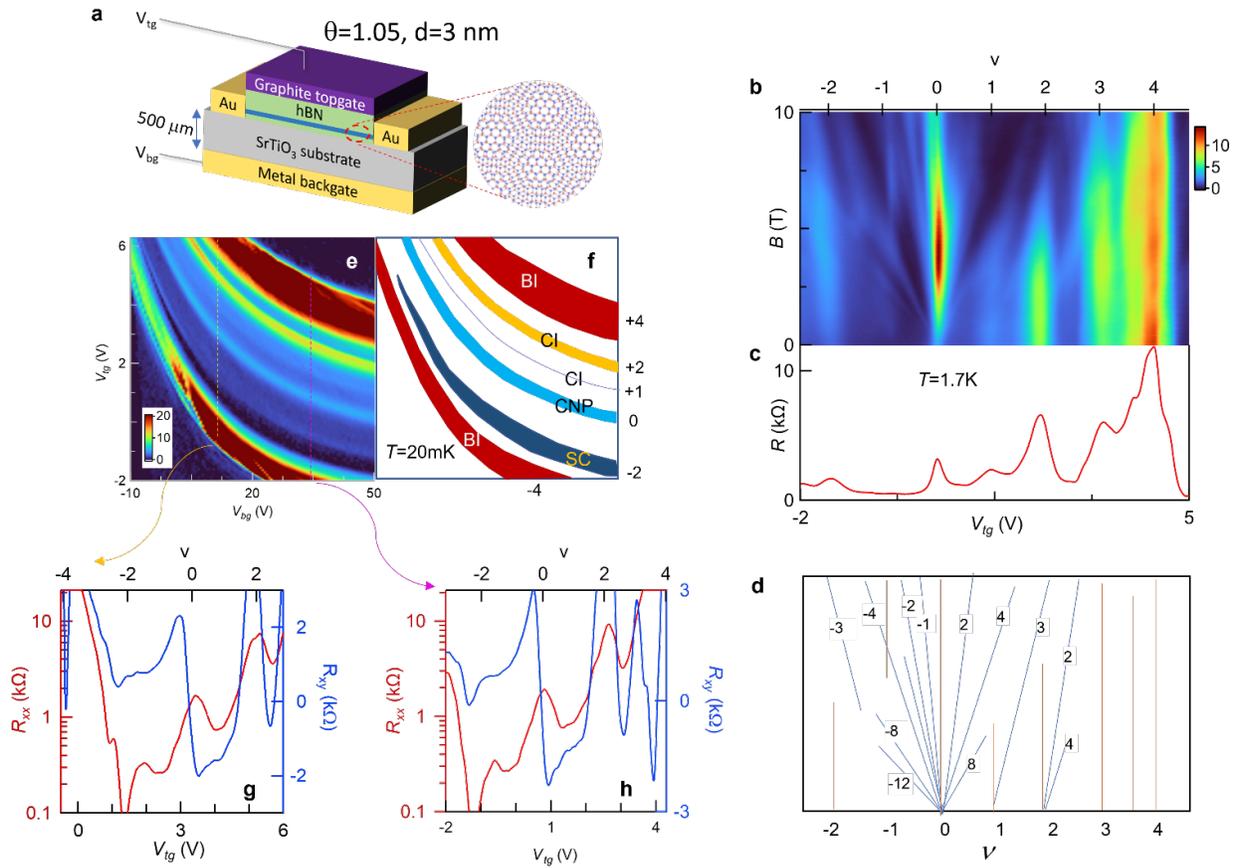

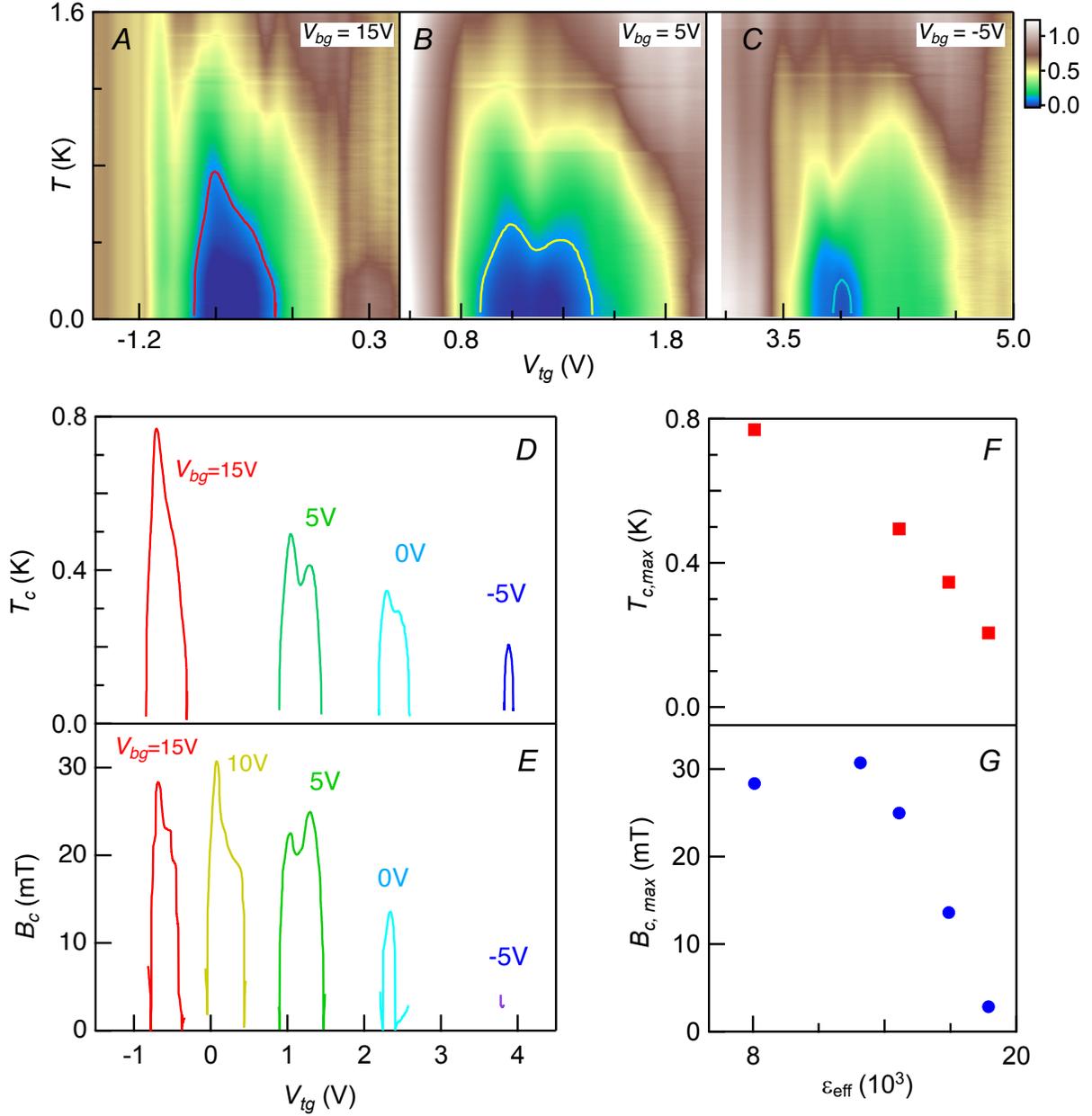

Fig. 2. Tuning superconductivity of D1 at $T$=20 mK. (A-C). Normalized resistance $R_{xx}/R_N$ vs $V_{tg}$ and $T$ at $V_{bg}$=15, 5, and -5V, respectively. The solid lines outline the superconducting dome, defined to be $R_{xx}/R_N$=20%. (D-E). $T_c$ and $B_c$ vs $V_{tg}$ at different $V_{bg}$ values, respectively. (F-G). Maximum $T_c$ and $B_c$ vs estimated effective dielectric constant of back gate.

Fig. 3. Transport data from Device D2 with $\theta=1.4°$ and $d=4.5$ nm at $T=20$ mK. (A). $R_{xx}(V_{tg}, B)$ and line trace $R_{xx}(V_{tg})$. (B). $R_{xx}(V_{tg}, V_{bg})$ phase diagram. (C). $dV/dI$ vs bias $I$ at $B=0$ (red) and $B=0.2$T (blue). (D). $dV/dI$ vs $I$ and $T$ at $B=0$. Data for (C-D) are taken at $V_{bg}=15$V and $V_{tg}=-2.23$ V. All color scales are in k$\Omega$.

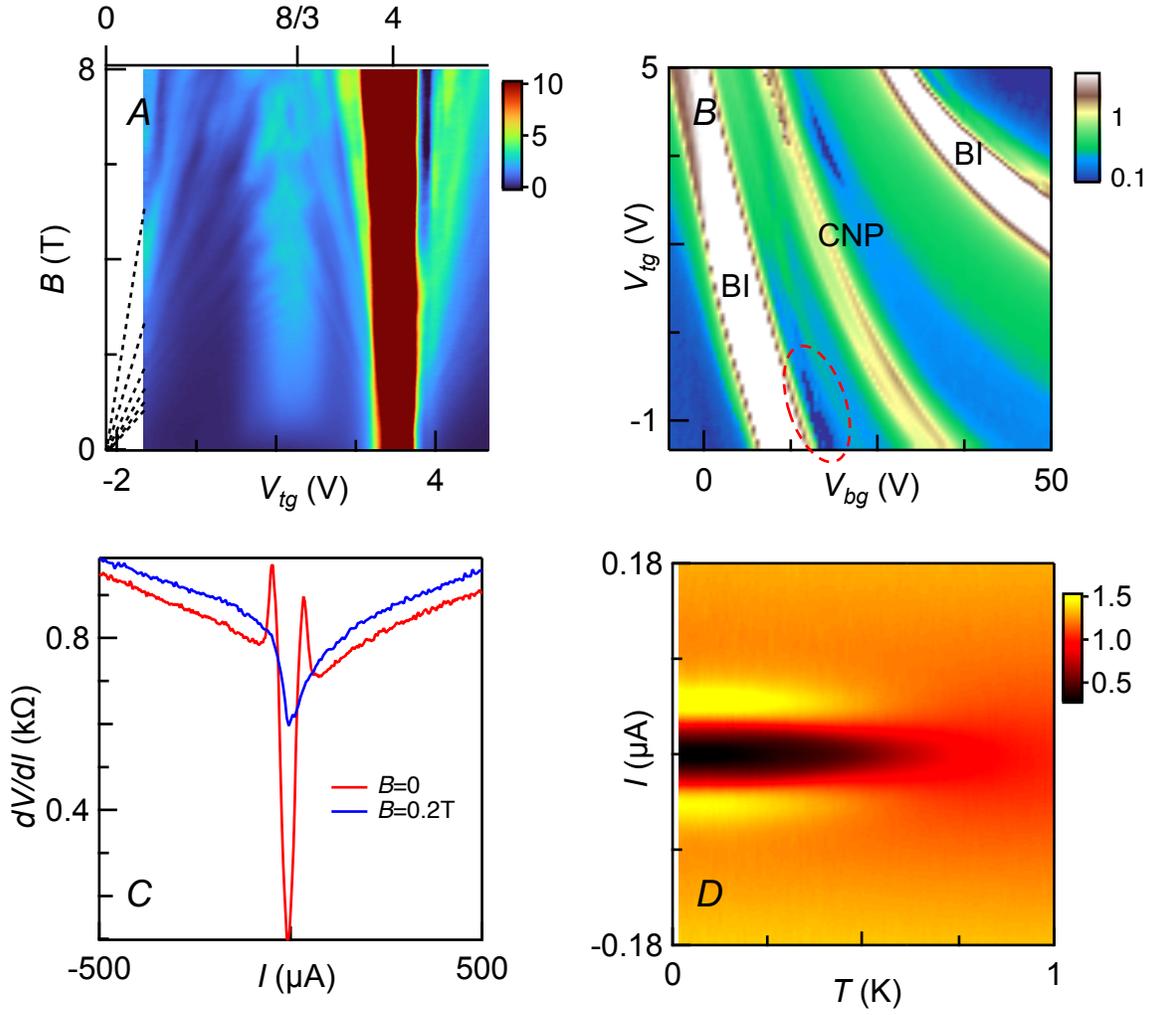

Fig. 4. Tuning the superconducting state in D2. (A). $T_c$ vs $V_{tg}$ at different $V_{bg}$ values, respectively. Here $T_c$ is taken as the temperature at which $R_{xx}/R_N = 50\%$. (B). Maximum $T_c$ vs estimated effective dielectric constant of back gate.

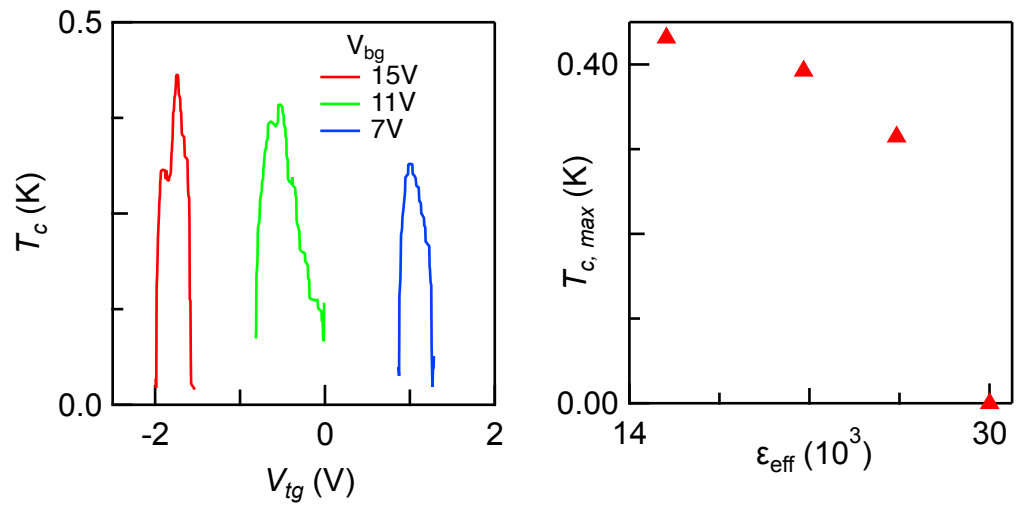